\documentstyle[12pt,epsfig]{article}

\begin{document}

\hfill{\bf {Preprint SNUTP-01/45}}
\vskip 2cm
\begin{center}
{\Large \bf A mechanism for the Double-Spin Asymmetry
in Electromagnetic $\rho$ Production at HERMES}

\vspace{3mm}

N.I. Kochelev,$^{1,2}$ D.-P. Min,$^3$ V. Vento,$^4$
and A.V. Vinnikov$^{3}$

{\small\it

$^1$ Bogoliubov Laboratory of Theoretical Physics,
     JINR, Dubna, Moscow region, 141980 Russia \\
$^2$ Institute of Physics and Technology, Almaty, 480082, Kazakhstan\\
$^3$ School of Physics and Center for Theoretical Physics,\\
     Seoul National University, Seoul 151-747, Korea \\
$^4$ Departament de F\'{\i}sica Te\`orica and Institut de F\'{\i}sica
     Corpuscular, Universitat de Val\`encia-CSIC E-46100,
     Burjassot (Valencia), Spain}

\end{center}

\begin{abstract}
We calculate the contribution of meson and pomeron exchanges to the
double-spin asymmetry in $\rho$-meson electromagnetic production
at HERMES energies. We  show that the
observed double-spin asymmetries, which are large, can be explained
by the interference between the
natural parity  $f_2$-secondary Reggeon
and  the unnatural parity anomalous
$f_1$ exchanges.
\end{abstract}

\section{Introduction}
The measurement of spin observables gives  very important information
on the structure of the strong interactions  \cite{review}.
Recently, the  HERMES Collaboration  has found a
significant ($\approx$ 20$\%$) double-spin asymmetry in elastic vector
meson electroproduction at energy $<E_\gamma>\approx 13$ GeV \cite{exp}.
This result is
quite intriguing since it was not expected within  models
of the vector meson production processes  based on convenient
mesonic and pomeron
exchanges.

There are two different approaches to describe the electromagnetic
production of light vector mesons  at intermediate  energies
\cite{friman} and \cite{laget}. The first  assumes that  only $\pi-$
and $\sigma-$ exchanges are relevant for explaining  the
 $\rho$ and $\omega$ cross sections and their relation.
The second  involves the pomeron, and the $\pi-$, $\sigma-$ and
 $f_2-$secondary reggeon exchanges. In  both of the approaches the main
contribution to the $\rho$ production cross section comes
from the natural
parity exchanges, e.g. pomeron, $\sigma$ and  $f_2$.

In spite of the fact that the $\pi NN$ coupling constant is very large,
the contribution of unnatural parity $\pi-$meson exchange to the elastic
cross-section is small and falls off rapidly with energy. At HERMES
energies its value is negligible in both models.

For spin observables the parity of the exchanges plays a crucial role.
For example, the double-spin asymmetry will not vanish only if there is
a considerable interference between the exchanges with natural and
unnatural parities.  It turns out, that the HERMES data are impossible to
describe by interference of the pion exchange with any existing natural
parity exchanges.

Recently we suggested a new unnatural parity anomalous $f_1$
trajectory,
with a very high intercept $\alpha_{f_1}\approx 1$ and a small
slope $\alpha_{f_1}\approx 0$, to explain  the
behavior of the elastic proton-proton, proton-antiproton and vector
meson photoproduction cross-sections at large energies and momentum transfers
\cite{our}. The connection of this new trajectory with spin physics has
been stressed, and its importance for the low $x-$ behavior of the nucleon
spin-dependent structure function $g_1(x,Q^2)$ and the double spin
asymmetries in diffractive reactions at large energies has been shown
\cite{our1}.

In here  we analyze the contributions of the $f_1$ exchange
to the double-spin asymmetries in $\rho$-meson
electromagnetic production  at intermediate energies
\footnote{The generalization of our model
to  $\omega$ and $\phi$-meson production is straightforward.}.

\section{The pomeron, $f_1$  and secondary Regge exchange contributions
to $\rho$-meson  production}

\begin{figure}[htb]
\centering
\epsfig{file=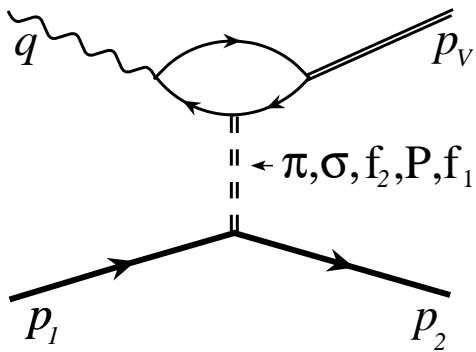,width=0.4\hsize}
\caption{The diagram describing $\rho$-meson electromagnetic production.}
\label{diag}
\end{figure}

The kinematics of the $\rho$-meson electromagnetic production off the
nucleon  is shown in Fig.\ref{diag}.
 The invariant variables of the
reaction are $s=(q+p_1)^2=(p_V+p_2)^2$ and $t=(q-p_V)^2=(p_1-p_2)^2$,
where $q$ and $p_1$ are the momenta of the initial photon and
proton, $p_V$ and $p_2$ are the momenta of the final $\rho$-meson and
proton.
At HERMES energies the mechanism for vector meson production is rather
complicated.
We will take into account the  contributions to the amplitude related to
the $t-$ channel
pomeron and the anomalous $f_1-$ exchanges \cite{our}, with a weak energy
dependence, together with the
secondary Regge $\pi-$, $\sigma-$ and  $f_2-$ exchanges,
which  have a rather strong energy dependence
\footnote{We will not include  the contribution
from the gluonic $G$-pole exchange related to
axial anomaly \cite{KV}. Its contribution is  important
at rather large $-t \geq 1 GeV^2$. HERMES only measures elastic vector
meson production cross section at low $-t\leq 0.6 GeV^2$.} .

For the pomeron exchange we use the Donnachie-Landshoff model
\cite{donland}.
The pomeron exchange amplitude is
\begin{eqnarray}
T^{P}= 12i m_V \beta_P^2
\sqrt{\frac{3 m_V \Gamma_{e^+ e^-}}
{\alpha_{em}}}.
\frac{(g_{\mu\nu} q^{\alpha} -g_{\nu,\alpha} p_V^{\mu}-g_{\mu\alpha}q^{\nu})
\varepsilon_{\gamma}^mu \varepsilon_V^{\nu}}
{q^2+t -m_V^2}
\bar u (p_2)\gamma_{\alpha} u(p_1) \times \nonumber \\ \times
\Bigl ( \frac{s}{s_P} \Bigr )^{\alpha_P (t)-1}exp\bigl \{ -\frac{i\pi}{2}
(\alpha_P (t) -1) \bigr \} F_P(t),
\label{pomampl}
\end{eqnarray}
where $m_V$ is the mass of the $\rho$ meson and  $\Gamma_{e^+ e^-}$ is
its leptonic width.
The quark-pomeron coupling $\beta_P$ can be fixed from fits \cite{landfit}
to total hadron-hadron cross sections. For the mass scale $s_P$ = 4
GeV$^2$ one obtains $\beta_P$ = 2.0 GeV$^{-1}$. The pomeron trajectory
is given by \cite{donland}
\begin{equation}
\alpha_P (t) = 1.08 +0.25 t.
\end{equation}
The form factor of the pomeron-NN vertex is \cite{donland}
\begin{equation}
F_P=\frac{4m_p^2-2.8t}{(4m_p^2-t)(1-t/0.71)^2}.
\end{equation}
It is also necessary to include an additional factor which takes into
account the
nonlocality of the pomeron vertex \cite{donland},
\begin{equation}
\frac{2\mu_0^2}{2\mu_0^2+m_V^2-t},
\end{equation}
where $\mu_0^2$ = 1.1 GeV$^2$.
This formula was obtained under the assumption that the effective
quark-quark scattering amplitude induced by the pomeron has the following form
\begin{equation}
T_{Pqq}=i\beta_P^2\bar q\gamma_\mu q\bar q\gamma_\mu q\Bigl ( \frac{s}{s_P} \Bigr )^{\alpha_P (t)-1}exp\bigl \{ -\frac{i\pi}{2}
(\alpha_P (t) -1) \bigr \},
\label{pom}
\end{equation}
which is supported by the Landshoff-Nachtmann two-gluon model of the
pomeron \cite{LN}.

Since at HERMES the energy is not very high, it is also necessary
to take into account the contribution from secondary reggeons.
The most important reggeons in the reaction under consideration
are the $\pi$, $\sigma$ and $f_2$-trajectories. The usual approach
to estimate  their contribution to vector meson production
is to fix their parameters from some low energy decay amplitudes
and  from fitting the total and differential elastic cross sections
for vector meson production
\cite{friman} and \cite{laget}. Therefore the predictive
power of such approaches is rather low.

To avoid this problem, we  calculate the contribution from secondary
Regge exchanges  using only information extracted from proton-proton and
proton-antiproton scattering in analogy to the  Donnachie-Landshoff
approach to the calculation of the pomeron contribution to
vector meson production
\cite{donland}. Within our approach one determines the parameters
of the secondary Regge exchanges between two quarks from the hadron-hadron
cross-sections. Then by using the assumption on a nonrelativistic shape
for the vector meson wave function, normalized to the leptonic decay
width  $\Gamma_{e^+ e^-}$, one  calculates  the contribution of any
exchanges to the vector meson production amplitude free of parameters.

Within this model,
using the following forms for effective $\pi$-quark and
$\sigma$-quark interactions
\begin{equation}
L_{\pi qq}=ig_{\pi qq}\bar q\gamma_5 q\pi^0, {\ }{\ }
L_{\sigma qq}=g_{\sigma qq}\bar qq\sigma,
\label{int}
\end{equation}
we obtain for the amplitudes
\begin{equation}
T^{\pi} = 4i\sqrt{\frac{m_{V}
\Gamma_{\rho\rightarrow e^+ e^-}}{3\alpha_{em}}}
\frac{g_{\pi qq}g_{\pi NN}
\epsilon_{\mu\nu\alpha\beta} q^{\alpha} p_V^{\beta}
\varepsilon_{\gamma}^{\mu} \varepsilon_{V}^{\nu}}
{(q^2+t -m_V^2)(t-m_{\pi}^2)}
\bar{u} (p_2)\gamma_5 u(p_1)
\label{pion}
\end{equation}
where
$g_{\pi qq} = 3/5 g_{\pi NN}$
is the value of the coupling of the $\pi^0$ meson
to quarks, and we use for the pion-nucleon coupling $g_{\pi NN}=13.28$
\cite{stoks}.
The $\pi$-meson form factor of the nucleon is given by \cite{piform}
\begin{equation}
F_{\pi}=\frac{\Lambda_\pi^2-m_{\pi}^2}{\Lambda_\pi^2-t},
\end{equation}
with $\Lambda=1.05 GeV$.
For the $\sigma$ exchange one has
\begin{eqnarray}
T^{\sigma}= 2g_{\sigma qq}g_{\sigma NN}
\sqrt{\frac{3 m_V \Gamma_{e^+ e^-}}{\alpha_{em}}}
\frac{\bigl ( 2 q_{\nu} p_{V\mu}-g_{\mu\nu}(m_{V}^2-t) \bigr ) }
{(q^2+t -m_{V}^2)(t-m_{\sigma})} \times \nonumber \\ \times
\varepsilon_{\gamma}^{\mu}\varepsilon_{V}^{\nu} \bar{u} (p_2) u(p_1)
R_{\sigma}(s,t)F_{\sigma}(t).
\label{sigma}
\end{eqnarray}
The parameters of the $\sigma$ exchange are taken from \cite{piform}:
$m_{\sigma}$=0.55 GeV,
$g_{\sigma qq}=1/3 g_{\sigma NN}$, $g_{\sigma NN}$ = 10.2,
\begin{equation}
F_{\sigma}(t)=\frac{\Lambda_\sigma^2-m_{\sigma}^2}
{\Lambda_\sigma^2-t},
\end{equation}
with $\Lambda_\sigma=2.0 GeV^2$.
In (\ref{pion}) and (\ref{sigma}) the  Regge
factor $R_{\pi,\sigma}(s,t)$ is \cite{guidal}
\begin{equation}
R(s,t)=\Bigl ( \frac{s}{s_0} \Bigr ) ^{\alpha(t)} \Gamma (1-\alpha(t))
\frac{1+e^{-i\pi\alpha(t)}}{2},
\label{regge}
\end{equation}
where $s_0=1 GeV^2$ and $\alpha(t)$ is the mesonic Regge trajectory
\begin{equation}
\alpha(t)_{\pi,\sigma}=\alpha'(t-m_{\pi,\sigma}^2),
\nonumber
\end{equation}
with slope $\alpha^\prime\approx  0.9 GeV^{-2}$.

We will assume with refs. \cite{laget} and \cite{landfit}  that  the
coupling of $f_2$ to the quarks has the same form as that of
the pomeron (\ref{pom}). Therefore the formula for the $f_2$  amplitude is
similar to  (\ref{pomampl}) changing the parameters to : $\beta_R$ =
3.5 GeV$^{-1}$, $s_0$ = 1 GeV$^2$, and the trajectory is now
$\alpha_R (t) = 0.55 +0.9 t$.
The values of  parameters  have been obtained from a fit
to the total hadron-hadron cross-sections
\cite{landfit}.

In a similar manner one can obtain the unnatural parity
 $f_1-$trajectory contribution to the
$\rho$-meson production amplitude.
The $f_1$ interaction with the quarks is
\begin{equation}
L_{f_1qq}=ig_{f_1qq}\bar q\gamma_\mu\gamma_5 q.
\nonumber
\end{equation}
By using this coupling we have for the amplitude,
\begin{eqnarray}
T^{f_1}= 4im_Vg_{f_1 qq}g_{f_1 NN}
\sqrt{\frac{3 m_V \Gamma_{e^+ e^-}}{\alpha_{em}}}
\frac{\epsilon_{\mu\nu\alpha\beta}q^{\beta} \varepsilon_{\gamma}^{\mu}
\varepsilon_V^{\nu} \bar{u} (p_2)\gamma_5 \gamma_{\delta} u(p_1)}
{(q^2+t -m_V^2)(t-m_{f_1}^2)} \times \nonumber \\ \times
(-g^{\alpha\delta}+(q-p_V)^{\alpha}(q-p_V)^{\delta}/m_{f_1}) F_{f_1}(t).
\label{f1}
\end{eqnarray}
In (\ref{f1}) the coupling of $f_1$ to the quarks,
$g_{f_1 qq} = g_{f_1 NN}$ = 2.5,  is fixed by using a constituent
quark model and the result of the analysis of the proton spin problem
within $f_1$ anomalous exchange model \cite{our}.
The form factor is given by
\begin{equation}
F_{f_1}(t)=\frac{1}{(1-t/m_{f_1}^2)^2},
\end{equation}
with $m_{f_1}=1.285 GeV$.

\section{The double-spin asymmetry}

\begin{figure}[htb]
\centering
\begin{minipage}[c]{0.4\hsize}
\epsfig{file=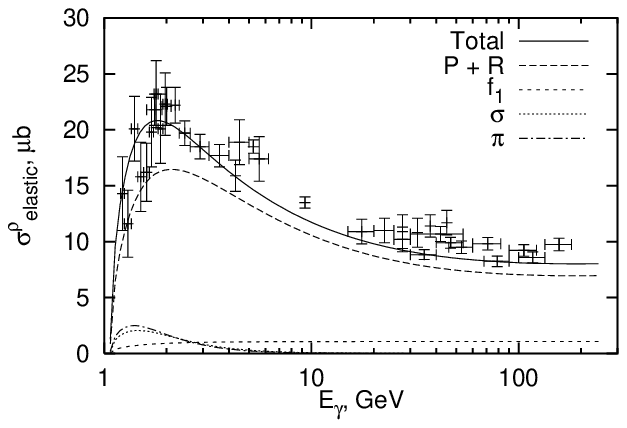,width=\hsize}
\caption{Contributions of hadronic exchanges to the total elastic
cross section of
$\rho$ meson photoproduction. The experimental data are taken from
\cite{crsexp}. \vspace {10.5mm} ~}
\label{crs}
\end{minipage}
\hspace*{0.5cm}
\begin{minipage}[c]{0.4\hsize}
\epsfig{file=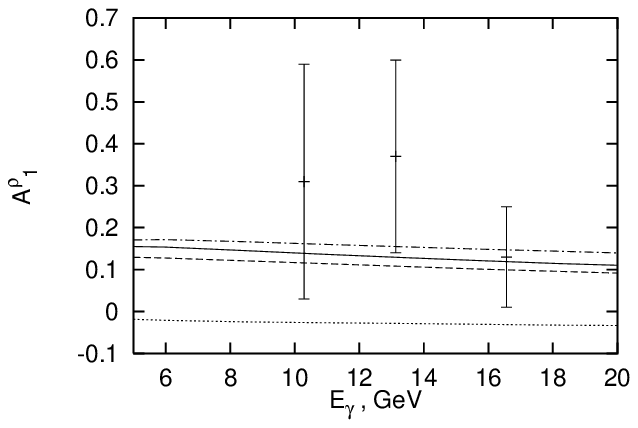,width=\hsize}
\caption{The calculated transverse double-spin asymmetry for $\rho$-meson
electromagnetic production compared with the experimental data from HERMES.
The solid line corresponds to pomeron-reggeon fit of
\cite{landfit};
the dashed line is for  the fit of ref.\cite{pdgfit}. The dotted line
shows the contribution
of the $f_1$-pomeron interference for the parameters from first fit
\cite{landfit}.}
\label{asymm}
\end{minipage}
\end{figure}

 The result of  the calculation of $\pi$-, $\sigma$-, $f_2-$, $f_1-$
contributions to the total elastic $\rho$ photoproduction
cross-section as a function of the photon energy
\begin{equation}
\sigma_{\pi,\sigma,f_2,f_1,P}
=\frac{1}{64\pi s|p_{cm}|^2}\int dt|M_{\pi,\sigma,f_2,f_1,P}|^2
\label{cross}
\end{equation}
are shown in Fig.(\ref{crs}). Our model describes the experimental data
well without tuning  the parameters. It is evident that for
HERMES kinematics,
$E_\gamma\approx 10\div 18 GeV$ \cite{exp}, the main contribution to the cross
section comes from the pomeron and $f_2$ exchange and the contributions
from $\pi$ and $\sigma$ exchanges are very small and therefore can be neglected.
The contribution of the $f_1$ exchange near the $\rho$ meson production
threshold
is smaller then the $\pi$ and $\sigma$ contributions.
But due to the very high intercept, the $f_1$ anomalous
trajectory,
has a non vanishing contribution at HERMES energies.
Due to its unnatural parity and negative signature,
 the corresponding amplitude
is proportional to the product of  helicities
of the photon and the proton.
Therefore the  interference of the $f_1$ amplitude with the natural parity
amplitudes  leads to
a nonzero double spin asymmetry,
\begin{equation}
A_1=\frac{\sigma_{1/2}-\sigma_{3/2}}{\sigma_{1/2}+\sigma_{3/2}},
\label{double}
\end{equation}
where $\sigma_{1/2,3/2}$ are the cross-sections for $\rho$-meson production
from different photon-proton helicity states.
Due to the large intercept $\alpha_{f_1}\approx 1$
 and negative signature,
the amplitude of the $f_1$ exchange is real with high
accuracy. Therefore one can expect that
 its interference with the nearly imaginary pomeron
amplitude is very small and its contribution to the double-spin asymmetry is
also small. At the same time the
$f_2$ amplitude contains
a significant real part and, as consequence, the main
contribution to the
double-spin  asymmetry should come from the interference
between the $f_2$ and the $f_1$ exchanges.

Within our model
all amplitudes have the same $Q^2$ dependence and therefore
the double-spin asymmetry does not depend on the value of $Q^2$.
In Fig.\ref{asymm} we compare  the contribution to the asymmetry
from the $f_1$-pomeron (dotted line), $f_1$-$f_2$ (dashed-dot line)
interferences  and the total result (solid line) with HERMES data
for an average  photon virtuality $<Q^2>$ = 1.7 GeV$^2$.
In the same figure we show the result
for a slightly different set of $f_2$ trajectory parameters
\cite{pdgfit}  by a dashed line.
As one can see, the contribution of the $f_1$-pomeron interference
is negative and rather small. The main contribution
comes, as it was to be expected, from the $f_2$-$f_1$ interference
and the total contribution explains the HERMES data within error bars.

\section{Conclusion}
We propose a new mechanism to explain the double-spin asymmetry
in $\rho$-meson
electromagnetic production off the nucleon at intermediate energies.
This mechanism is related to the existence of a new anomalous
unnatural parity, negative signature $f_1$ Regge trajectory
which was introduced
recently in \cite{our} to explain features of diffractive reactions at large
momentum transfers  and the anomalous behavior of the spin-dependent
structure function $g_1(x,Q^2)$.
The interference of the $f_1$ exchange with the natural parity
$f_2$ secondary reggeon exchange
gives the very large contribution to the double spin asymmetry.
The calculated value for the total asymmetry is
$A_1\approx 10-15\% $, which is compatible
with the measured value 24$\pm$11{\%}.
For $\phi$- and $J/\Psi-$ mesons the main contribution to the
double-spin asymmetry should come from interference of
pomeron and the $f_1$ trajectory, because the $f_2$ couplings
with these mesons vanish in OZI limit.
Therefore one can expect small  and
negative double-spin asymmetries for both mesons.

We have shown that the $f_1$ anomalous exchange gives the
biggest contribution to the spin-dependent cross sections
of vector meson production, not
only at very large energies \cite{our1},
but also in intermediate energy range.
\section*{Acknowledgements}
We are grateful to  A.Borissov and K.Lipka for
useful discussion.
This work was partially  supported by SEUI-BFM2001-0262 , RFBR-01-02-16431,
INTAS-2000-366 grants and BK21 and Heisenberg-Landau programs.
The work of AV, NK and DPM is also supported in part by KOSEF 199-2-111-005-5
and KRF 2001-015-DP0085.


\begin{thebibliography}{99}
\bibitem{review} R.L.Jaffe, hep-ph/0101280.
\bibitem{exp}
A.~Airapetian {\it et al.}  [HERMES Collaboration],
Phys.\ Lett.\  {\bf B513}, 301 (2001).
\bibitem{friman}
B.~Friman and M.~Soyeur,
Nucl.\ Phys.\  {\bf A600}, 477 (1996).
Y.~Oh, A.~I.~Titov and T.~S.~Lee, nucl-th/0004055.
\bibitem{laget}
J.~M.~Laget and R.~Mendez-Galain,
Nucl.\ Phys.\  {\bf A581}, 397 (1995), \\
J.~M.~Laget,
Phys.\ Lett.\ {\bf B489}, 313 (2000).
\bibitem{our}
N.~I.~Kochelev, D.~P.~Min, Y.~Oh, V.~Vento and A.~V.~Vinnikov,
Phys.\ Rev.\ {\bf D61}, 094008 (2000).
\bibitem{our1}
Y.~Oh, N.~I.~Kochelev, D.~P.~Min, V.~Vento and A.~V.~Vinnikov,
Phys.\ Rev.\  {\bf D62}, 017504 (2000).
\bibitem{guidal}
M.~Guidal, J.~M.~Laget and M.~Vanderhaeghen,
Nucl.\ Phys.\ {\bf A627}, 645 (1997).
\bibitem{KV}
N.~I.~Kochelev and V.~Vento,
Phys.\ Lett.\ {\bf B515}, 375 (2001);\\
 hep-ph/0110268.
\bibitem{stoks}
V.~G.~Stoks, R.~Timmermans and J.~J.~de Swart,
Phys.\ Rev.\  {\bf C47}, 512 (1993).
\bibitem{piform}
R.~Machleidt,
Adv.\ Nucl.\ Phys.\  {\bf 19}, 189 (1989).
%\bibitem{f1form}
%A.~Liesenfeld {\it et al.}  [A1 Collaboration],
%Phys.\ Lett.\  {\bf B468}, 20 (1999).
\bibitem{donland}
A.~Donnachie and P.~V.~Landshoff,
Phys.\ Lett.\ {\bf B185}, 403 (1987).
\bibitem{landfit}
A.~Donnachie and P.~V.~Landshoff,
Phys.\ Lett.\  {\bf B296}, 227 (1992)
\bibitem{LN}  P.~V.~Landshoff and O.~Nachtmann,
Z.\ Phys.\ {\bf C35} (1987) 405.
\bibitem{pdgfit}
D.~E.~Groom {\it et al.}  [Particle Data Group Collaboration],
Eur.\ Phys.\ J.\  {\bf C15}, 231 (2000).
\bibitem{crsexp}
W.~Struczinski {\it et al.}  [Aachen-Hamburg-Heidelberg-Munich
                  Collaboration],
Nucl.\ Phys.\  {\bf B108}, 45 (1976), \\
Y.~Eisenberg {\it et al.},
Nucl.\ Phys.\  {\bf B42}, 349 (1972), \\
J.~Breitweg {\it et al.}  [ZEUS Collaboration],
Eur.\ Phys.\ J.\  {\bf B2}, 247 (1998);\\
Y.~A.~Aleksandrov {\it et al.},
Yad.\ Fiz.\  {\bf 32}, 651 (1980); \\
R.~M.~Egloff {\it et al.},
Phys.\ Rev.\ Lett.\  {\bf 43}, 657 (1979); \\
J.~Barth {\it et al.}  [SAPHIR Collaboration],
Acta Phys.\ Polon.\  {\bf B29}, 3321 (1998).

\end{thebibliography}
\end{document}